# Quantifying the Origins of Life on a Planetary Scale

Caleb Scharf[*], Leroy Cronin[**]

[*]*Columbia Astrobiology Center, Columbia Astrophysics Laboratory, 550 West 120th Street, New York, NY 10027, USA, caleb@astro.columbia.edu*
[**]*School of Chemistry, University of Glasgow, Glasgow, G12 8QQ, UK, lee.cronin@glasgow.ac.uk*

**Abstract:** A simple, heuristic formula with parallels to the Drake Equation is introduced to help focus discussion on open questions for the origins of life in a planetary context. This approach indicates a number of areas where quantitative progress can be made on parameter estimation for determining origins of life probabilities. We also suggest that the probability of origin of life events can be dramatically increased on planets with parallel chemistries that can undergo the development of complexity, and in solar systems where more than one planet is available for chemical evolution, and where efficient impact ejecta exchange occurs, increasing the effective chemical search space and available time.

**Main text:** The question of whether or not life exists beyond the confines of the Earth environment is intimately related to the question of life's origins on Earth or anywhere else. For example, if just one instance of life with an independent origin is detected - whether in the solar system or on a suitable exoplanet - estimates of the Bayesian probability for life across the universe will be significantly improved. Specifically, the rate of abiogenesis per solar system could, in principle, be constrained to be at least 1 event per Gyr.*(1)* Equally, if a detailed mechanism for the terrestrial origins of life (OoL) were identified and tested, a complementary estimate of the rate of planetary abiogenesis should be possible. Further, the architecture of a given solar system might also lead to widely different probabilities e.g. the proximity of Mars and Earth and their potential exchange of material might result in a fundamentally larger search space in which OoL events could occur. Discovering complex organic molecules on Mars could help elucidate this possibility.*(2)*

At present OoL science involves a wide range of approaches, theories, and opinions. A considerable body of work now exists on hypothetical early terrestrial life and its precursor chemistry. Including the RNA-world hypothesis, prior polymerization chemistry, autocatalytic processes and metabolic origins, as well as the presumed transition to fully cellular organisms. Significant work has also been undertaken on inorganic 'template' chemistry, the planetary evolution of suitable material precursors, and important substrates for the emergence of a biology - including the study of electrochemical gradients in hydrothermal vents systems and processes such as serpentization as energy sources. In addition, the genomic exploration of the modern Earth's microbial life is providing ever-deeper insight to the nature and evolution of the biosphere's nested metabolic processes,*(3)* which may provide clues to the original energy landscape of pre-biotic chemical cycles. Theoretical and computational work on emergent and dynamical systems, complexity, and 'dry' (not wet lab based) artificial life (A-Life) are also contributing to our picture of the fundamental principles of biological system operations.

However, there is no strong consensus opinion that supports any single OoL hypothesis or timeline - whether for the specific OoL events on the Earth or for any plausible alternatives. It is therefore impossible to construct a first principles quantitative estimate of an abiogenesis probability for the young Earth. Nonetheless, abiogenesis did clearly occur at least once, and we do possess general information about Earth's composition and a number of notional histories for the surface and near-surface environment during the planet's first few hundred million years following the solar system's initial collapse and condensation from the pre-solar nebula. In this short paper a simple formula (§2) is proposed to encourage further consideration of the issues at play in linking OoL (or abiogenesis) to planetary environments. This approach parallels similar descriptive and pedagogical methods applied in assessing the likelihood of extraterrestrial communications in the Search for Extraterrestrial Intelligence (SETI). A simple application of this formula is made and the results discussed in §3. In §4 a number of modifications and extensions are considered that should help bring focus to the regions of parameter space where tractable progress can be made on building a genuinely quantitative evaluation of OoL likelihoods.*(4)*

In 1961 Frank Drake introduced an equation to illustrate the factors involved in estimating the potential number of communicative civilizations in the Milky Way galaxy. This formula, now generally known as The Drake Equation, has served as a useful tool for focusing discussion on the extent to which we do, or don't, have constraints on its various factors, and for stimulating ideas about how to make scientific progress on the problem of finding life in the universe over a given period of time.*(5)* There are however serious limitations to actually using the Drake formula to produce meaningful estimates of life's frequency in the universe. A particular issue is that although the formula seeks to describe the entire galaxy's living population it does not explicitly allow for the possibility of life spreading and expanding between planets. Thus, the factors in the equation may *not* be independent at all.

Here we propose that a focus on the parameters involved in planetary OoL offers a better constrained entry point to this type of estimation, and could produce practical results, especially when combined with the process of seeking potentially 'habitable' exoplanetary systems. Assuming that life on Earth did originate *in-situ* - i.e. that the transition, or the key transitions, between a non-living state and a living state occurred within the planetary environment and not from far off-world (e.g. panspermia beyond the planetary system) - it should be possible to construct a high-level quantitative description of the basic features of OoL, irrespective of the details. Such a description will necessarily make a quantitative connection to planetary properties, such as mass, element abundance and energy resources.

The following OoL frequency equation is therefore proposed for $N_{abiogenesis}$, the number of abiogenesis or origin events within time $t$,*(6)* where the definition of such events and the associated factors are discussed in the subsections below:

$$N_{abiogenesis}(t) = N_b \cdot \frac{1}{\langle n_o \rangle} \cdot f_c \cdot P_a \cdot t$$

### $N_b$: number of potential building blocks
Modern terrestrial life can be deconstructed into a large, but finite, set of functional chemical components. For example, the major building blocks are often cited as consisting of families of carbohydrates, lipids, proteins, and nucleic acids. Additional blocks might include metal ions as part of the metallome (e.g. coordinated zinc ions stabilizing protein fold structures, iron in hemoglobin, copper in hemocyanin and other metalloproteins), silicon (e.g. in plants) and many trace elements.

However, life today does not necessarily reflect all the details of early biochemistry, or earlier complex chemistry that was able to self-sustain routes to competing biologies that ultimately led to biology as we know it. This is because the entire process is, arguably, more likely to reflect the consequence of billions of years of evolution of a series of earlier chemical toolkits, all with decreasing complexity and therefore, increasing probability of emergence over a short period of time. In that context it might be advisable to break these components down further and consider the raw elemental (atomic) abundances as the first factor in the OoL equation. An additional advantage in doing so is that it simplifies the connection of $N_b$ to bulk planetary properties, thereby linking this expression back to the scope of the Drake Equation and our direct physical knowledge of the Earth and other planets.

In this case $N_b$ represents a *maximal* set of building blocks for life that can be estimated from (see below) the total mass and composition of the outer planetary layers. In the Earth's case it is self-evident that not all of the atoms of Earth's outer layers are incorporated into living states, nor are all available for that at any given time. To account for this the factor $f_c$ is described further below. Also, any collection of building blocks must be able to interact with each other and provide routes by which a subset of building blocks can drive the assembly and complexification of a subset of new building blocks -that have improved function, structure, and robustness. Chemical cycles, operating on these building blocks, will provide raw materials as well as facilitate the assembly of more sophisticated building blocks from the minimal set, but these cycles are fundamentally dependent upon the input building block set.

Chemical cycles could also allow the same building blocks, in different sequences, to make polymers and other infrastructure that can allow both control of assembly and maintenance of sequence and expression of function. Also, the kinetic connections between the different cycles and sequences constitutes a minimal kinetic-genetic mechanism by which a more robust and ultimately evolvable genetic machinery may develop, and this might even utilize an network of so-called 'inorganic' reactions to start developing function.*(7)*

**$\langle n_o \rangle$ : mean number of building blocks per 'organism', or biochemically significant system.**
The definition of 'organism' in this context is broad and is related to a quantitative definition of life - the precise details of which remain an open question at this time. For OoL it could be taken to mean a minimal life-form, one that is capable of homeostasis, reproduction, and open-ended evolution. However, we might determine that such a life-form exists either as an encapsulated, cellular system or as a distributed, non-local, but interdependent and autocatalytic chemical system. Similarly, a biochemically significant system could be construed to be a system or environment that is a direct and necessary precursor to a minimal life-form, with an extremely high probability of the transition to that life-form. Properly evaluating $\langle n_o \rangle$ may therefore require the development of a coherent, quantitative, definition of the 'aliveness' of a collection of building blocks.

**$f_c$: fractional availability of building blocks during time $t$.**
Within any given timespan, $t$, it can be assumed that only a certain fraction of the total number of potential building blocks in a planetary environment will actually be available for life. Availability in this context may be defined through a number of factors: 'free' (unbound to other molecular or atomic species), 'mobile' (capable of physical transport, not restricted except in terms of necessary localization), or 'energetic' (energetically favored chemical bonding and incorporation into a system). These factors can be strongly dependent on planetary details, such as environmental temperature. The use of this factor (i.e. not incorporating it directly into $N_b$) allows the weighting given to intrinsic planetary properties such as crust/mantle ratios or global elemental abundances to be separated and adjusted independently. As we discuss below (§4) $f_c$ can be significantly expanded to include explicit treatments of planetary habitability and the interplay between living systems and environment - e.g. the sequestration of building blocks by prior life.

**$P_a$ : probability of assembly per unit time.**

This factor refers to the probability of an abiogenesis 'event' per unit time through the assembly of the necessary building blocks. The necessary definition of an event is of course very poorly understood. Abiogenesis could be highly spatially localized, within a free-floating vesicular structure for example, or in an inorganic substrate's intimate topography. Alternatively, abiogenesis could refer to a non-localized, sequential, series of assembly steps or processes that lead directly and irreversibly to an evolvable system. We purposefully treat $P_a$ as a catch-all that circumvents the need to go into the (unknown) mechanical details.

**Evaluating $N_{abiogenesis}$ and $P_a$.**

As described in §1 and §2 above, we propose the OoL frequency expression (Equation 1) to help facilitate and focus discussion on OoL science in the same way that the Drake Equation is applied to the quest for life elsewhere in the universe. It is not meant to be genuinely predictive, nor complete in the form given here. Despite all of these shortcomings, it is instructive to evaluate a range of plausible values for the factors, given the observation that $N_{abiogenesis} \geq 1$ for the Earth over a likely interval of ~100 Myr from ~4 Gya.

In Table 1 a set of crude but plausible values are listed for the equation's factors. The basis is a simplistic overview of the Earth. Potential building blocks are taken to be the elemental (atomic) constituents of the upper planetary environment (crustal mass, oceans and atmosphere). This is very likely a gross over-estimate, but even a few orders of magnitude variation will not alter the subsequent conclusions below. The number of building blocks required for abiogenesis (see above) is derived from an approximate mean value for the number of atoms in a bacterium. The fractional availability of building blocks is derived from estimates of the modern ratio of Earth's total biomass to the mass of the upper planetary environment. And $t$ is assumed to be the first 100 Myr following the final major episode of planetary assembly.

Table 1: Plausible values for the factors in the OoL frequency equation for the Earth.

| Factor | Value | Basis for value |
| --- | --- | --- |
| $N_b$ | ~$10^{49}$ | Number of atoms in Earth's modern crust + ocean + atmosphere |
| $\langle n_o \rangle$ | ~$10^{11}$ | Number of atoms in bacterium (dry weight) |

| $f_c$ | ~$10^{-14}$ | Fraction based on the ratio of modern Earth's total estimated biomass to the total mass of crust + ocean + atmosphere |
| $P_a$ | ~$10^{-32}$ | Assuming $N_{abiogenesis} = 1$ within a 100 Myr interval |

Although the chosen values are spectacularly approximate, the linear nature of the equation ensures that variations in any factors by even few orders of magnitude do little to alter the basic outcome. In this scenario the abiogenesis probability per unit time is found to be $P_a \sim 10^{-32}$. Thus, in this case the equation indicates that a very small assembly probability per unit time could in principle be compensated for by the large scale of a planetary 'search engine' - readily producing at least one abiogenesis event within a geologically appropriate timespan.

This statement is independent of fine details. It does not, for example, hinge on any analysis of the combinations and permutations of atoms required to form a functional biochemical system or the odds of assembly computed on that basis, all of that complexity is folded implicitly into $P_a$.

In setting $N_{abiogenesis}$(100 Myr)=1, we implicitly assume that after time $t$ there is a sole, successful abiogenesis event. Perhaps that is incorrect. For example, if $P_a$ is actually much larger, there would either be many independent abiogenesis events that are indistinguishable in our current model for the history of life on Earth (leading to the inaccurate conclusion that $N_{abiogenesis} = 1$), or many independent abiogenesis events that undergo rapid selection competition, or simple chemical failure, to winnow the population to a single progenitor model for all of life today.

Other games can be played. For example, one can demand that $P_a = 1$ in a 100 Myr timespan (i.e. $P_a = 10^{-8}$ in a year) *and* that there is a single abiogenesis outcome, $N_{abiogenesis} = 1$. Retaining the Table 1 values for $N_b$ and $f_c$, a literal interpretation would be that $\langle n_o \rangle \sim 10^{43}$. In other words, an inevitable net abiogenesis in 100 Myr (via a global process of interaction and chemical selection over geological timescales) might involve ~0.0001% of the potential building blocks on a planet like the Earth. Again, we emphasize that in this initial simple, heuristic, form the equation is not meant to be rigorous. It can however drive discussion of the 'macro' and 'micro' scale issues for OoL and form the basis for a more complete quantitative measure of abiogenesis rates, as described below.

The above formula (Equation 1) assumes that all factors are independent, and that they contribute linearly to $N_{abiogenesis}$. This, of course, may not be true. For example, as building blocks (atoms or simple molecules) are incorporated into pre-biotic molecular structures their availability will change. If polymerization is a key part of abiogenesis this abundance dependency could be important as an asymptotic property. Similarly, $P_a$ is unlikely to be a single, time-independent variable. Rather $P_a$ will have a complex dependency on many factors, from the scale of the planetary environment (and therefore $N_b$), to the *relative* abundance of building blocks (i.e. a block dependent $f_c$), and the chemical pathways involved in assembly.

The factor $f_c$ in particular can be readily expanded out. For example:

$$f_c = f_p \cdot f_a \cdot f_e \cdot (1 - f_l)$$

where $f_p$ is the fraction of the total planetary environment (containing $N_b$) within the "habitable zone" (e.g. using the common astronomical definition of temperature and vapor-pressure such that open water remains liquid), $f_a$ is the fraction of building blocks in solution or part of a bio-accessible substrate, $f_e$ is the fraction of building blocks with access to the necessary energy (chemical, thermal, photonic) to drive assembly to a biologically meaningful assembly, and $f_l$ is the fraction of building blocks that may already be incorporated into abiogenesis related assemblies (i.e. not available at time $t$ for incorporation in any new abiogenesis events).

In this instance, $f_p$ is an explicit function of the planetary climate state, which is in turn determined by properties such as the planetary spin-orbit state, the parent star, and the planetary atmospheric composition. $f_a$ is more complex but could be estimated from initial planetary veneer compositions. The factor $f_e$ might also be estimated from models of initial planetary properties. In the simplest case, for $t = 0$, the factor $f_l$ is assumed to be 0. Allowing $f_l > 0$ introduces explicit non-linearity in the equation since we might naively expect $f_l = (N_l \cdot \langle n_o \rangle)/N_b$, where $N_l$ is the *number* of existing 'organisms' at time $t$ that are sequestering building blocks away from abiogenesis events (assuming they don't actually contribute positively to further abiogenesis). Thus, the full expression for abiogenesis events now becomes:

$$N_{abiogenesis}(t) = N_b \cdot \frac{1}{\langle n_o \rangle} \cdot f_p \cdot f_a \cdot f_e \cdot \left(1 - \frac{N_l \langle n_o \rangle}{N_b}\right) \cdot P_a \cdot t$$

As Equation 3 is written, there is no explicit incorporation of the concept of dilution. A planet may represent a large set of $N_b$ and high $f_c$ as defined, but the building blocks could still be highly diluted, for example by large water bodies. Indeed, the dilution problem of an ocean-covered Hadean/Archean Earth is well appreciated (where monomer concentration in open oceans appears too low for polymerizations to occur). In this case, since $N_b$ presumably scales with planetary size, any potential dilution factor may also scale with $N_b$. Thus there must be a more complex behavior of $f_c$, where the 'availability' of building blocks should include a statistical factor to account for localized concentration or spatial clustering (e.g. hydrothermal vent 'oasis' environments, tidal zones).

The final $N_{abiogenesis}$ after time $t$ could also depend on a survival or failure rate. Similar to the comments in §3, $P_a$ could be larger, but balanced by a probability of extinction due to molecular fragility, system fragility, competition, or just a steady rate of random failure. It is possible that material exchange between terrestrial-type planetary surfaces could serve to greatly amplify the chemical 'search space' within a planetary system. If we assume, for example, that the Hadean Earth and Noachian Mars provided chemical incubators that were distinct from other solar system locales (e.g. primordial cometary or asteroidal environments) and well suited to the emergence of increasing pre-biotic chemical complexity, then exchange between these bodies could have accelerated the exploration of chemical novelty.

There is good evidence that the architecture and dynamical evolution of the solar system resulted in early periods of significant impact events on both Earth and Mars, and subsequent ejection and exchange of lithospheric material.*(8)* Although the exchange of viable organisms between these planets is still very much an open question,(9) the exchange of molecular species may be much more likely. Furthermore, the energy-scale of impact events required to launch material onto interplanetary trajectories need not be so destructive as to 'reset' indigenous pre-biotic molecular complexity and populations on a planetary scale.

In this scenario $N_{abiogenesis}$ becomes coupled across planets in a potentially complicated, but important way. The variation in $N_b$ and $f_c$ due to impact ejecta exchange may be very small, however changes in $P_a$ as defined may be significant. As described above (§2.4) we see $P_a$ as the probability of an 'event' - either a specific chemical assembly or a non-localized sequence or chemical system that leads irreversibly to an evolvable system. Thus, in the same spirit as Equation 1, we suggest that the effect of inter-planetary exchange could be most simply modeled by assuming that all receipts of material (e.g. Earth receiving Mars material) result in an *increase* in $P_a$. In other words, it is assumed that 'new' molecular species are brought to the planet and that these species positively influence the pathway to abiogenesis.

For example, this could behave as a simple geometric series:

$$P_a' = P_a \cdot E^n$$

where $E$ is the factor by which $P_a$ is increased by a single exchange, $n$ is the total number of 'receipt' events due to impact ejecta (e.g. the number of impact ejecta creating collisions that occur on Mars), and $P_a'$ is the revised assembly event probability.

The factor $E$ is of course unknown, and likely varies with each exchange. However, in an extreme scenario, each planetary incubator generates an entirely unique set of molecular species in between impact exchange events. Very crudely speaking, in this case it would be easy for $E \sim 2$ and, trivially, if $n \sim 100$ then $P_a' \sim 10^{30} P_a$. Thus, the original value of $P_a$ estimated for an isolated Earth in Table 1 (assuming $N_{abiogenesis} = 1$ in 100 Myr) would be modified and would approach unity.

The value of $n$ can in principle be constrained in the solar system by data on impact histories and our models of solar system dynamical evolution. For the Earth, evidence suggests that the Late Heavy Bombardment (thought to occur between 3.8 and 4.1 Gya) may have involved more than 20,000 Earth impacts capable of launching ejecta to interplanetary space.(10) Mars appears likely to have experienced a bombardment that was at least as intense.(11) Thus, the likelihood of a large $n$ for the Earth is high. That also suggests that $E$ could be very modest yet the final $P_a'$ might still approach unity. We therefore suggest that, in an extreme case, the natural exchange of chemical toolkits between just two young terrestrial-type planets, could have an enormous influence on the overall likelihood of abiogenesis in a planetary system. It could in fact represent the difference between life arising or not.

The life that we know of on Earth is united by a common biochemical alphabet. In the above discussion, the atoms or molecules referred to as building blocks are expected to become functional members of an 'assembly' event. This is implicit, for example, in the way that $\langle n_o \rangle$ is used. However, we cannot yet rule out the possibility of alternate, or 'shadow' life operating with a different alphabet (e.g. an inorganic one). We also cannot not rule out the possibility that abiogenesis occurred on Earth *because* of interactions between mutually exclusive building blocks that together acted to dynamically increase the evolvable information content of the pre-biosphere. In other words, it is possible that the building blocks of life today are the product of a combination of actions of earlier sets of building blocks that could not 'cross-assemble' but that could, together, generate the necessary complexity and selection for organic abiogenesis. Equation 1 makes no assumptions about $P_a$, it is simply the probability of assembly per unit time when all other necessary conditions are met, but a further exploration of the potential influence of mutually exclusive building blocks could prove instructive.

In summary we have sketched out a macro-scale, or top-down, perspective on OoL questions through the formulation of a simple, but instructive equation for the frequency of abiogenesis in a planetary environment. Given the current explosive progress in the discovery and characterization of exo-planets, the factors that we suggest play a role in OoL likelihoods could motivate specific observational data (e.g. the configuration and material exchange fluxes of planets per system, and their chemical composition / heterogeneity). We propose that this approach may be useful for stimulating future discussions, but should also have a genuinely quantitative role in studying the origins of life.

*Acknowledgments*: This work stems from discussions held at the Earth Life Science Institute (ELSI) at the Tokyo Institute of Technology, as part of the ELSI Origins Network (EON) road-mapping workshop "A Strategy for Origins of Life Research". CS thanks EON for financial support and hosting at ELSI made possible by a grant from the John Templeton Foundation. CS also acknowledges the support of Columbia University's Research Initiatives in Science and Engineering during the course of this work. LC thanks the EPSRC for financial support through grants EP/L023652/1 and EP/J015156. LC also acknowledges the support of the University of Glasgow.

**References**

1. D. S. Spiegel, E. L. Turner, *Proc. Natl. Acad. Sci. USA,* **109**, 395 (2012).

2. S. A. Benner, K. G. Devine, L. N. Matveeva, D. H. Powell *Proc. Natl. Acad. Sci. USA*, **97**, 2425 (2000).

3. Falkowski, P. G., Fenchel, T., DeLong, E. F., 2008. The Microbial Engines That Drive Earth's Biogeochemical Cycles. Science, 320, 1034-1039

4. C. Maccone, *Ori. Life. Evo. Bio.,* **41**, 609 (2011).

5. M. M. Cirkovic, *Astrobiology*, **4**, 225 (2004).

6. C. H. Lineweaver, T. M. Davis, *Astrobiology*, 2, 293 (2001).

7. A. R. de la Oliva, V. Sans, H. N. Miras, J. Yan, H. Zang, C. J. Richmond, D. –L. Long, *Angew. Chem. Int. Ed.*, **51**, 12759 (2012).

8. B. J. Gladman, J. A. Burns, M. Duncan, P. Lee, & H. F. Levison, *Science*, **271**, 1387 (1996).

9. C. Meyer, J. Fritz, M. Misgaiski, et al., *Meteoritics and Planetary Science*, **46**, 701 (2011).

10. G. Ryder, C. Koeberl, S. J. Mojzsis, Heavy Bombardment on the Earth at ~3.85 Ga: The Search for Petrographic and Geochemical Evidence. Origin of the Earth and Moon, 475 Tucson: University of Arizona Press (2000).

11. D. M. Burt, L. P. Knauth, & K. H. Wohletz, The Late Heavy Bombardment: Possible Influence on Mars. LPI Contributions, 1439, 23 (2008).